\documentclass[prl,twocolumn,showpacs]{revtex4}
\usepackage{graphicx}

\renewcommand{\v}[1]{{\bf #1}}

\newcommand{\w}{{\omega}}

\def\eqa{\begin{eqnarray}}
\def\eea{\end{eqnarray}}
\newcommand{\eq}{\begin{equation}}
\newcommand{\ee}{\end{equation}}

\newcommand{\<}{\langle}
\renewcommand{\>}{\rangle}

\renewcommand{\Im}{{\rm Im}}

\newcommand{\ua}{\uparrow}
\newcommand{\da}{\downarrow}

\newcommand{\Del}{\Delta}

\newcommand{\si}{\sigma}

\newcommand{\cG}{ {\cal G} }

\newcommand{\cN}{ {\cal N} }

\begin{document}

\title{Impurity resonance states in electron-doped high $T_c$ superconductors}
\author{Yuan Wan}
\author{Hai-Dong L\"{u}}
\author{Hong-Yan Lu}
\author{Qiang-Hua Wang}
\email[E-mail address:]{qhwang@nju.edu.cn}
\address{National Laboratory of Solid State Microstructures \& Department of Physics, Nanjing University,
Nanjing 210093, China}


\begin{abstract}
Two scenarios, i.e., the anisotropic s-wave pairing (the s-wave
scenario) and the d-wave pairing coexisting with
antiferromagnetism (the coexisting scenario) have been introduced
to understand some of seemingly s-wave like behaviors in electron
doped cuprates. We considered the electronic structure in the
presence of a nonmagnetic impurity in the coexistence scenario. We
found that even if the AF order opens a full gap in quasi-particle
excitation spectra, the mid-gap resonant peaks in local density of
states (LDoS) around an impurity can still be observed in the
presence of a d-wave pairing gap. The features of the impurity
states in the coexisting phase are markedly different from the
pure AF or pure d-wave pairing phases, showing the unique role of
the coexisting AF and d-wave pairing orders. On the other hand, it
is known that in the pure s-wave case no mid-gap states can be
induced by a nonmagnetic impurity. Therefore we proposed that the
response to a nonmagnetic impurity can be used to differentiate
the two scenarios.
\end{abstract}

\pacs{PACS numbers:74.25.Jb,74.20.-z,73.43.Jn}

\maketitle

\section{I. Introduction}
Up to now, the pairing symmetry of electron-doped high $T_c$
superconductor is still under debate, and various interpretations
for experimental results are controversial. On one hand, the phase
sensitive Josephson junction experiments indicate d-wave pairing
symmetry pairing,\cite{Tsuei00} and meanwhile angle-resolved
photoemission spectra (ARPES) \cite{Sato01} and electronic Raman
spectra\cite{Blumberg02} suggest a non-monotonic d-wave energy gap
as a function of momentum. On the other hand, tunnelling
spectra\cite{Kashiwaya98} and low temperature specific
heat\cite{Balci04} show more or less s-wave like behavior, i.e.
the absence of zero bias conductance peak in the spectrum. Two
scenarios have been introduced to explain these results. One is
the s-wave scenario, in which the s-wave character is regarded to
be intrinsic, but it is incompatible with the phase-sensitive
Josephson junction experiment. The other one is the coexisting
scenario in which the coexisting antiferromagnetic (AF) order
disguises the d-wave character of superconductivity
(SC).\cite{Yuan04,Luo05} The idea is as follows. First of all, the
in-plane AF order has been observed by neutron
scattering\cite{Kang03} and transport experiments\cite{Lavrov04},
and this AF order should be considered naturally. Further, since
the AF order opens a full gap in single particle excitation
spectrum, which combines with the nodal d-wave SC gap to form an
effective full single-particle gap. In a recent paper it is shown
that a more careful analysis of the Raman spectra could
disentangle the effective gap into its AF and pairing
components.\cite{Lu06} It remains to see whether a single-particle
probe, such as the scanning tunnelling microscopy (STM) could be
used to identify the ingredients of the effective gap and thence
differentiate the two scenarios.

In fact, impurity in unconventional superconductors proves to be a
useful tool to characterize various SC orders. For example, in a
d-wave superconductor a zero energy impurity resonant state
appears as a hallmark of the d-wave pairing symmetry.
\cite{Balatsky95}. On the contrary, in conventional s-wave
superconductor the resonant states lie at the gap edge, which is
known as Yu-Shiba-Rusinov state.\cite{Yu65} The drastic difference
is due to the phase structure of the two SC orders: in the d-wave
case the phase of pairing wave function changes sign across the
nodal lines, while in s-wave case such no sign change occurs.
Mid-gap impurity resonant states can also occur in $p_x+ip_y$ and
$d_x+id_y$ superconductors.\cite{Wang04} Impurity resonant states
can also be used to characterize the electronic structure in some
materials.\cite{Wehling06} Furthermore, it is also proposed that
by inspecting the line-shape of the resonant peak as a function of
temperature around a nonmagnetic strong impurity can differentiate
the phase disorder scenario and d-density wave scenario of the
pseudo-gap phase in hole-doped superconductors.\cite{Wang02}
Therefore, STM measurements of the impurity states can provide
important messages on the underlying system.\cite{Fisher07} The
question we now ask is if AF and d-wave pairing coexist, what is
the nature of the impurity state, and in particular whether a low
energy resonance state can still arise around the impurity.

In this paper we calculate the LDoS around a nonmagnetic impurity
in electron-doped high $T_c$ superconductor. Our main results are
as follows. Firstly, although the AF order gaps the
quasi-particles and the bulk density of states is s-wave like, two
mid-gap impurity resonant states, lying symmetrically at positive
and negative energy, can be observed in the presence of a d-wave
SC order. Secondly, the two resonant peaks in LDoS approach and
cross each other when the impurity scattering strength increases
up to the unitary limit. At an intermediate scattering strength
the two peaks merge into one peak at the Fermi energy. Since such
a mid-gap impurity resonant state does not exist in the s-wave
scenario, the two scenarios for the electron-doped superconductor
are differentiable by STM measurements of the nature of the low
energy impurity states. The structure of the rest of the paper is
as follows. In section II we describe the model and method. In
section III we present the results. First we obtained the phase
diagram of the system in the absence of impurities. Second, we
discuss analytically a toy model with particle-hole symmetry to
show the existence of impurity resonance states in the coexisting
scenario. Third, we switch back to the actual situation in the
phase diagram, and calculate the LDoS in the presence of a
nonmagnetic impurity. In order to see the particular role of the
coexistence orders, we compare the actual case to the cases with
AF or d-wave pairing alone. Section IV is a summary of the work.

\section{II. Model and method}
We adopt the t-t'-t''-J model on a square lattice with the
hamiltonian,
\begin{eqnarray}
H =
-t\sum_{\langle{}ij\rangle{}_{1}\sigma}(\tilde{c}^{\dagger}_{i\sigma}\tilde{c}_{j\sigma}+h.c.)
-t'\sum_{\langle{}ij\rangle{}_{2}\sigma}(\tilde{c}^{\dagger}_{i\sigma}\tilde{c}_{j\sigma}+h.c.)\nonumber\\
-t''\sum_{\langle{}ij\rangle{}_{3}\sigma}(\tilde{c}^{\dagger}_{i\sigma}\tilde{c}_{j\sigma}+h.c.)
+J\sum_{\langle{}ij\rangle{}_{1}}\v S_{i}\cdot{}\v S_{j}
\end{eqnarray}
Here $\tilde{c}_{i\sigma}$ and $\tilde{c}^{\dagger}_{i\sigma}$ are
Fermion operators subject the the non-double-occupancy constraint,
$\langle{}ij\rangle{}_{1}$, $\langle{}ij\rangle{}_{2}$, and
$\langle{}ij\rangle{}_{3}$ denote the first, second and third
nearest neighbor pairs respectively. It should be noted that we
work in the hole picture, so that a hole in the above model
represents a physical electron-double-occupancy in electron doped
cuprates. To cope with the above t-J model, we insist in calling
the electron-double-occupancy as holon unless indicated otherwise.
For the parameters we choose $|t|$ as the unit of energy, so that
$t=-1$, $t'=0.32$, $t''=-0.16$ and spin exchange integral
$J=0.3$.\cite{Yuan04} We emphasize that the choice of parameters
are conventional, but our results are not sensitivity to the
parameters.

We apply the slave boson mean field theory (SBMFT), within which
the projected Fermion operator is decoupled to a spinon and a
holon part $\tilde{c}_{i\sigma}=h^{\dagger}_{i}f_{i\sigma}$, and
the restriction of no double occupancy is replaced by the
constraint
$h^{\dagger}_{i}h_{i}+\sum_{\sigma}f^{\dagger}_{i\sigma}f_{i\sigma}=1$.\cite{Ubbens92}
In the mean field theory this operator constraint is replaced by
its average counterpart. The holons are assumed to condense at
zero temperature, so that
$h_{i},h^{\dagger}_{i}\rightarrow{}\sqrt{x}$, where $x$ is the
doping level. The spin-exchange term is decoupled in a standard
way into hopping, pairing and spin-moment channels,
\cite{Brinckmann01}
\begin{eqnarray}
\v S_{i}\cdot\v S_j\rightarrow{}-3/8(\langle\hat{\chi}_{ij}^{\dagger}\rangle\hat{\chi}_{ij}+h.c.)\nonumber\\
-3/8(\langle\hat{\Delta}_{ij}^{\dagger}\rangle\hat{\Delta}_{ij}+h.c.)+(\langle{}\hat{m}_{i}\rangle{}\hat{m}_{j}+\hat{m}_{i}\langle{}m_{j}\rangle{})
\end{eqnarray}
Here the bracket $\langle\cdot \rangle$ denotes mean value of an
operator.
$\hat{\chi}_{ij}=\sum_{\sigma}f^{\dagger}_{i\sigma}f_{j\sigma}$ is
the hopping operator,
$\hat{\Delta}_{ij}=f_{i\uparrow}f_{j\downarrow}-f_{i\downarrow}f_{j\uparrow}$
is the singlet paring operator, and
$\hat{m}_{i}=\sum_{\sigma}\sigma{}f^{\dagger}_{i\sigma}f_{i\sigma}$
is the magnetic moment operator. The mean value of these operators
are the corresponding order parameters.

We introduce 4-spinors,\eqa \Psi_{k}=(f_{\v
k\uparrow},f^{\dagger}_{-\v k\downarrow}, f_{\v k+\v
Q\uparrow},f^{\dagger}_{-\v k-\v Q\downarrow})^{T}\nonumber\eea in
the momentum space. Here $\v Q=(\pi,\pi)$ is the nesting vector.
The mean field Hamiltonian can be written in terms of the
4-spinors as $H^{f}_{MF}=\bar{\Psi}_{\v k}h_{\v k}\Psi_{\v k}$,
where the $4\times 4$ matrix $h_{\v k}$ is given by
\begin{equation}\label{bloch}
h_{\v k}=\left(\begin{array}{cccc}
\xi_{\v k} & \Delta_{\v k} & 2Jm & 0 \\
\Delta_{\v k} & -\xi_{\v k} & 0 & 2Jm \\
2Jm & 0 & \xi_{\v k+\v Q} & \Delta_{\v k+\v Q} \\
0 & 2Jm & \Delta_{\v k+\v Q} & -\xi_{\v k+\v Q}
\end{array} \right),
\end{equation}
where $\xi_{\v k}$ is the mean-field dispersion of the Fermions,
$\Delta_{\v k}$ is the paring gap function in momentum space,
which are given by,
\begin{eqnarray*}
\xi_{\v k}&&=-\mu-(2xt+3/4\chi)[\cos(k_x)+\cos(k_y)]\\
&& -4xt'\cos(k_x)\cos(k_y)-2xt''[\cos(2k_x)+\cos(2k_y)],\\
\Delta_{\v k}&&=-3/4J\Delta[\cos(k_x)+\cos(k_y)].
\end{eqnarray*}
Here $m=\<\hat{m}\>$, $\chi=\<\hat{\chi}\>$,
$\Delta=\<\hat{\Del}\>$, and $\mu$ is the chemical potential to
fix the doping level. All of these parameters are obtained through
self-consistent mean field calculation. The spinon's propagator in
the above representation is obtained as $\cG_{\v
k}(i\omega_{n})=(i\omega_{n}\v I-h_{\v k})^{-1}$. The coherent
part is given by $\cG_{coh}=x\cG$ upon convoluting with the holon
part.

For later discussion we define a $2\times 2$ Green's function
$G_0(\v r_i,\v r_j
;\tau)=-\hat{T}\<(f_{i\ua},f_{i\da}^\dagger)^T(\tau)(f_{j\ua}^\dagger,f_{j\da})(0)\>$
for the 2-spinors $(f_{i\ua},f_{i\da}^\dagger)^T$ in the real
space and $\tau$ is the imaginary time. It's Fourier component in
frequency $\w_n$ is related to $\cG$ as follows,
\begin{eqnarray}
G_{0}(\v r_i,\v r_j;i\w_n)&&=\frac{1}{N}\sum_{\v k\in{}MBZ}e^{i\v
k\cdot (\v r_i-\v r_j)}[\cG_{\v k,I}+\cG_{\v
k,II}e^{i\v Q\cdot \v r_j} \nonumber\\
&& +\cG_{\v k, III}e^{i\v Q\cdot \v r_i}+\cG_{\v k,IV}e^{i\v
Q\cdot (\v r_i-\v r_j)}],
\end{eqnarray}
where the summation is over the magnetic Brillouin zone (MBZ) and
$N$ is the size of lattice. The subscripts $I,II,III,IV$ indicate
the left-top,right-top,left-bottom and right-bottom $2\times 2$
block elements of $\cG$.

The next step is to obtain the Green's function in the presence of
a single impurity. We model the impurity by a single-site
potential located at the origin, and adopt the T-matrix
formulation\cite{Balatsky06,Wang02,Wang04} to calculate the
perturbed Green's function $G$ by \eqa G(\v r_i,\v r_j;z)=G_0(\v
r_i,\v r_j;z) +G_0(\v r_i,0;z)T(z)G_0(0,\v r_j;z),\eea where \eqa
T(z)=&& V\sigma_3+V\sigma_3G_0(0,0;z)T(z).\eea Here $z$ is the
complex continuation of $i\w_n$, $V$ is the nonmagnetic impurity
potential, and $\si_3$ is the third Pauli matrix. In the T-matrix
formulation the correction of mean-field order parameter induced
by impurity is ignored. This was shown to be sufficiently
consistent with a full self-consistent calculation.\cite{Franz96}

Finally, the LDoS can be measured by STM directly, which is given
by \eqa \cN(\v r;\omega)=-\frac{1}{\pi}\Im[G_{11}(\v r,\v
r;-\omega+i\eta)+G_{22}(\v r,\v r;\omega+i\eta)],\label{LDoS}\eea
where $\eta$ denotes an infinitely small positive number. The
convention of the sign before $\w$ is chosen for the
electron-doped case under our concern.

\section{III. Results}

\subsection{A. SBMFT phase diagram}

\begin{figure}
\includegraphics[width=0.45\textwidth]{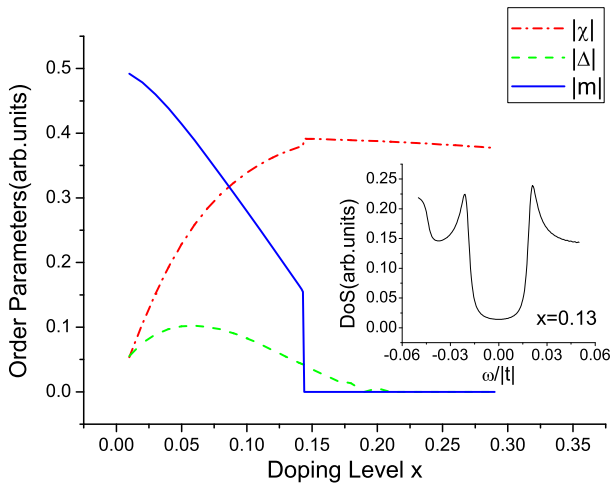}
\caption{\label{phasediagram}(Color online) SBMFT phase diagram of
electron doped superconductor. The red dash-dotted line, green
dashed line and blue solid line denote the hopping magnitude,
d-wave paring order and staggered magnetic moment respectively.
The inset shows the bulk density of states at $x=0.13$.}
\end{figure}

We first present the main results of our SBMFT calculation. The
mean field phase diagram is shown in Fig.\ref{phasediagram}. The
calculation is done on a $1000\times 1000$ lattice. We can see
that there are two phase transition points. At $x\approx 0.14$
there is a first order phase transition form the AF phase to the
paramagnetic phase. On the other hand, at $x\approx 0.20$ there is
a second-order phase transition from the superconducting phase to
the normal phase. Thus in the range $x<0.14$ the AF and SC order
coexist. The optimal doping level lies at $x\approx 0.06$.

Given the mean field order parameter as above, we can calculate
the bulk density of states (DoS) in different phases. In the
following discussion we will concentrate on the coexisting phase.
A typical result at doping level $x=0.13$ is shown as the inset of
Fig.\ref{phasediagram}. The U-shaped DoS indicates the absence of
node in single particle excitation spectrum, and have been
observed in point contact tunnelling spectra but was interpreted
as the character of s-wave pairing.\cite{Kashiwaya98} In our case,
it is a result of coexisting AF and SC orders. Similar behavior
was found by Yuan \textit{et al} using t-U-V model.\cite{Yuan04}

\subsection{B. Impurity states in a particle-hole symmetrical case}

\begin{figure}
\includegraphics[width=0.45\textwidth]{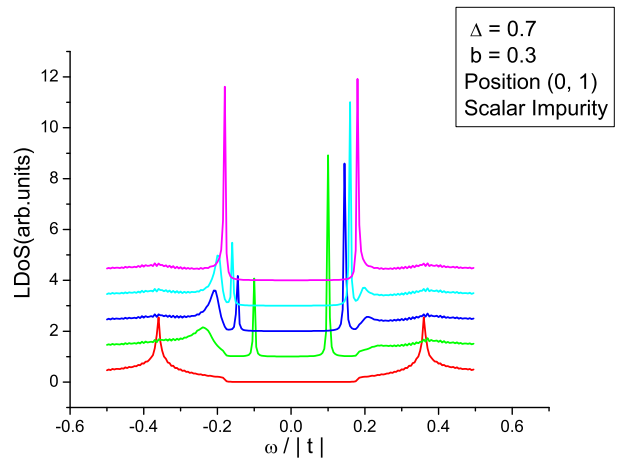}
\caption{\label{partiholesym}(Color online) The LDoS at position
(0,1) for various impurity potential strengths in a particle-hole
symmetrical system. Form bottom to top the potential strengths are
successively 0, 5, 10, 15, and 1000, in units of $|t|$. Vertical
offset is used for clarity.}
\end{figure}

To see whether resonant impurity states could appear in the
coexisting phase, we digress to consider a the particle-hole
symmetrical case that is analytically tractable. To reach this
case we simply set $\mu=0$, $t'=0$, and $t''=0$. We have
$\Delta_{\v k}=-\Delta_{\v k+\v Q}$ and $\xi_{\v k}=-\xi_{\v k+\v
Q}$. The unperturbed on-site Green's function $G_0(0,0;z)$ can be
calculated explicitly as,
\begin{equation}
G_0(0,0;z)=\frac{1}{(2\pi)^2}\int_{BZ}d\v k^2\frac{z-b}
{z^2-\xi^2_\v k-\Delta^2_{\v k}-b^2}\times\sigma_0
\end{equation}
Here the $\sigma_0$ is the two by two unit matrix, and $b=2Jm$ is
the Curie-Wess potential. The T-matrix is given by
$T^{-1}(z)=V^{-1}-G^0(0,0;z)$. The position of the resonant state
is given by $\det{T^{-1}}=0$. In the unitary limit $V\to\infty$
the resonance condition is given by
$\det{G_0(0,0;z)}=0$.\cite{Wang04} Since the off-diagonal elements
of $G_0$ is zero due to the d-wave pairing symmetry, the resonance
occurs at $z=b$. Since the $11$ ($22$) component of $G$ describes
particles (holes), we see that another resonance should occur at
$z=-b$ in the particle picture alone. This is also evident from
the two contributions in the expression for $\cN$ in
Eq.(\ref{LDoS}).

If the impurity potential is finite, the situation is more
complicated. In Fig.\ref{partiholesym} we present the dependence
of LDoS on potential strength. In the calculation we simply set
$\chi=0$, $\Delta=0.7$ and $b=0.3$ for illustration. The LDoS is
for site (0,1), a nearest neighbor of the impurity site. When the
potential strength is $V/t=5$, there appears already two resonant
peaks lying symmetrically at the positive and negative sides
already. With increasing potential strength the resonant peaks
shift toward the gap edges, and the height of peaks is strongly
enhanced. Similar phenomenon has been observed in the study of
organic superconductivity with bond-dimmerization.\cite{Tanuma03}
In the next subsection we discuss how the resonance behaves in the
realistic situation.

\subsection{C. Impurity states in electron doped cuprates}

\begin{figure}
\includegraphics[width=0.45\textwidth]{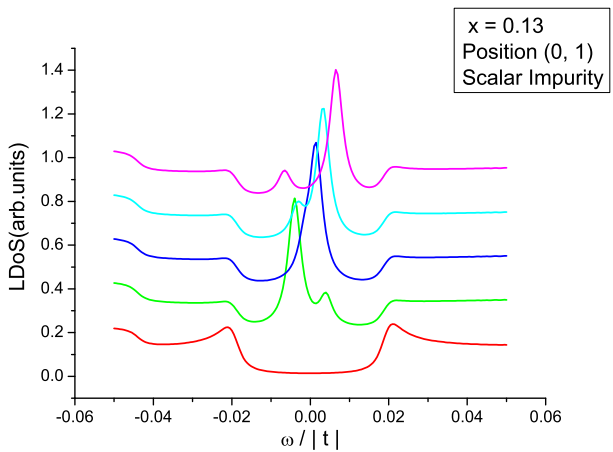}
\caption{\label{actual}(Color online) The same plot as
Fig.\ref{partiholesym} but for the actual cases described by the
phase diagram.}
\end{figure}

In this subsection we discuss the realistic situation in electron
doped cuprates. We choose the result of SBMFT as the input of
T-matrix formulation. The evolution of LDoS at site (0,1) with
impurity potential strength at the doping level $x=0.13$ is shown
in Fig.\ref{actual}. The mean field order parameters are
$\chi=-0.37$, $\Delta=0.054$, and $m=0.19$. At first we note that
in the unitary limit $V/|t|=1000$, there are two mid-gap resonant
states lying symmetrically at the positive and negative energy,
but the height of the two peaks are different due to the effect of
coherence factor in $G^0(\v r,0;z)$, due to the particle-hole
asymmetry in the present case. The two peaks are visible for
$V/|t|\geq 5$. With increasing potential strength, we see that the
two peaks seem to cross each other. For $V/|t|=10$ the two peaks
meet and merge into one peak which lies right at the Fermi energy.
It is therefore clear that resonance states do appear in the
coexisting phase, even though the nodal d-wave gap is disrupt by
the AF gap.

\subsection{D. Pure AF order and pure SC order case}

\begin{figure}
\includegraphics[width=0.45\textwidth]{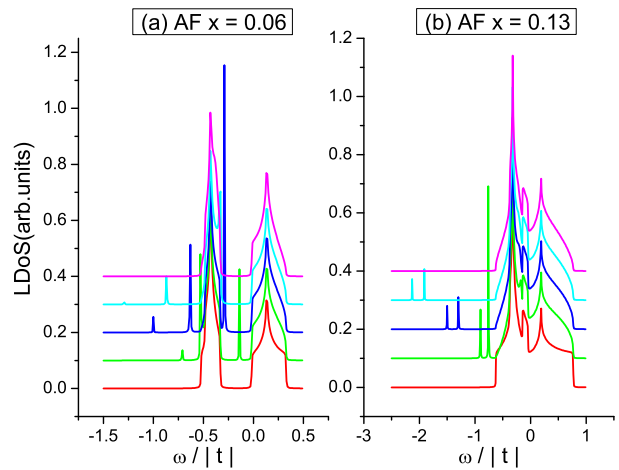}
\caption{\label{AFM}(Color online) The same plot as
Fig.\ref{partiholesym} but in pure AF cases at two doping levels.}
\end{figure}

\begin{figure}
\includegraphics[width=0.45\textwidth]{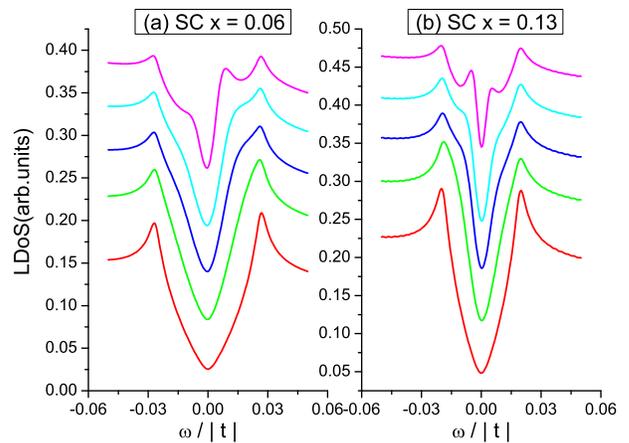}
\caption{\label{SC}(Color online) The same plot as
Fig.\ref{partiholesym} but in pure d wave SC cases at two doping
levels.}
\end{figure}

In order to understand the unique role of the coexisting orders,
we compare the results in the pure d-wave case and the pure AF
order case. We set the d-wave SC order parameter to zero in
(\ref{bloch}) and leave the AF order parameter unchanged for pure
AF order case. Conversely, we set the AF order to zero for pure
d-wave case. In principle one should do the self-consistent
calculation again in each case. We have done so but found no
significant changes to the remaining order parameters.

First, we consider the pure AF case. The result is shown in
Fig.\ref{AFM}(a) and (b) for $x=0.06$ and $x=0.13$, respectively.
For $x=0.06$, the AF order is strong enough to split the two
bands, and impurity resonant peaks lie in the gap as well as
outside the band. For $x=0.13$, the AF order is weak and the two
AF-order-induced bands overlap, and then impurity resonant states
(the sharp peaks ) lie outside bands. The positions of resonant
peaks move toward higher energy for higher potential strengths,
and are pushed to infinity in the unitary limit.

Second we turn to the pure d-wave case shown in Fig.\ref{SC}(a)
and (b) for $x=0.06$ and $x=0.13$. We can identify two resonant
peaks at the two doping levels in the unitary limit, but the
strong particle-hole asymmetry has weaken the strength of peaks.
On the other hand, the two peaks never cross each other. Moreover,
for weak potential strength these peaks are indiscernible.

To close this section, we conclude by comparison that two
well-defined mid-gap resonant peaks appear and cross each other
with increasing impurity potential only in the coexisting phase
with both AF and d-wave SC order.

\section{IV. Summary}
In this work we discuss the impurity resonant state in the
coexisting phase with d-wave SC and AF orders. We demonstrate
analytically the existence of mid-gap resonance states in a
putative particle-hole symmetrical case, and present the results
in the realistic case, where the resonance peaks can shift and
switch with increasing potential strength. At an intermediate
potential strength, the two peaks merge at the Fermi level. These
unique features do not appear in the separate pure AF or pure
d-wave pairing states. It is also know that mid-gap impurity
states do not appear in a pure $s$-wave phase. Thus, the impurity
state can be used to differentiate the s-wave scenario and
coexisting scenario.

\acknowledgments{This work was supported by NSFC 10325416, the Fok
Ying Tung Education Foundation No.91009, the Ministry of Science
and Technology of China (under the Grant No. 2006CB921802 and
2006CB601002) and the 111 Project (under the Grant No. B07026).}

\end{document}